\pdfoutput=1

\documentclass[aip, amsmath, amssymb, reprint, nofootinbib]{revtex4-1}

\usepackage[pdftex]{color}              
\usepackage[pdftex]{graphicx}
\usepackage[pdftex,colorlinks=true,linkcolor=black,pdfview=FitH,pdfstartview=FitH]{hyperref}
\hypersetup{citecolor=black}
\usepackage{amsfonts}
\usepackage{amsmath}
\usepackage{bm}                 % bold math

\newcommand{\sslabel}[1]{\label{sec:#1}}
\newcommand{\elabel}[1]{\label{eqn:#1}}
\newcommand{\flabel}[1]{\label{fig:#1}}
\newcommand{\tlabel}[1]{\label{tab:#1}}

\newcommand{\cref}[1]{\ref{cpt:#1}}
\newcommand{\sref}[1]{\ref{sec:#1}}
\newcommand{\eref}[1]{\ref{eqn:#1}}
\newcommand{\fref}[1]{\ref{fig:#1}}
\newcommand{\tref}[1]{\ref{tab:#1}}

\newcommand{\MEsec}[2]{\section{\sslabel{#2}#1}}

\newcommand{\listBegin}{\begin{tabular}{cp{4.5in}}}
\newcommand{\listEnd}{\end{tabular}}

\newcommand{\matrixBegin}[1]{\left[\!\!\left[ \begin{array}{#1}}
\newcommand{\matrixEnd}{\end{array} \right]\!\!\right]}

\newcommand{\beq}[1]{\begin{equation}\elabel{#1}}
\newcommand{\eeq}{\end{equation}}

\newcommand{\beqa}[1]{\begin{eqnarray}\elabel{#1}}
\newcommand{\eeqa}{\end{eqnarray}}

\newcommand{\MEfig}[4]{\begin{figure}[#1] \includegraphics[width=#2\textwidth]{#3} \vspace{-5pt}
\caption{\flabel{#3}#4}\end{figure}}

\newcommand{\abs}[1]{\left| #1 \right|}

\newcommand{\set}[1]{\left\{ #1 \right\}}

\newcommand{\fun}[2]{\,{#1}\!\left( {#2} \right)}

\newcommand{\tfun}[2]{\fun{\textrm{#1}}{#2}}

\usepackage{siunitx}

\DeclareSIUnit\rtHz{$\sqrt{\si{\Hz}}$}
\DeclareSIUnit\mrtHz{\meter\per\rtHz}

\definecolor{spring}{rgb}{0.7,0.9,0.7}
\definecolor{brick}{rgb}{0.7,0.2,0.1}
\definecolor{redHL}{rgb}{1.0,0.5,0.5}

\def\gw{gravitational wave}
\def\gws{gravitational waves}
\def\GW{Gravitational Wave}

\newcommand{\smallMatrix}[1]{\begin{pmatrix}#1\end{pmatrix}}

\def\Fout{\mathbf{\bar{o}}}
\def\Fsig{\mathbf{\bar{s}}}
\def\Fsifo{\Fsig_0}
\def\Fhd{\mathbf{\bar{b}}_{\zeta}}

\newcommand{\FinN}[1]{\mathbf{\bar{i}_{#1}}}
\def\Fin{\FinN{n}}

\newcommand{\TnN}[1]{\mathbf{T_{#1}}}
\def\Tn{\TnN{n}}
\def\Tfc{\TnN{fc}}
\def\Tifo{\TnN{ifo}}

\def\TwoP{A_2} % audio-SB to 2-photon matrix

\def\Kifo{\mathcal{K}}
\def\Ksr{\mathcal{K}_{sr}}

\graphicspath{{pic/}{plot/}}

\begin{document}

%%%%%%%%%%%%%%%%%%%%%%%%%%%%%%%%%%%%%%%%
% Header

%%%%%%%%%%%%%%%%%%%%
\title{Realistic Filter Cavities for Advanced \GW\ Detectors}
\author{M. Evans}\affiliation{Massachusetts Institute of Technology, Cambridge, Massachusetts 02139, USA}
\author{L. Barsotti}\affiliation{Massachusetts Institute of Technology, Cambridge, Massachusetts 02139, USA}
\author{J. Harms}\affiliation{INFN, Sezione di Firenze, Sesto Fiorentino, Italy}
\author{P. Kwee}\affiliation{Massachusetts Institute of Technology, Cambridge, Massachusetts 02139, USA}
\author{H. Miao}\affiliation{California Institute of Technology, Pasadena, California 91125, USA}

%%%%%%%%%%%%%%%%%%%%
\begin{abstract}
The ongoing global effort to detect \gws\ continues to push the limits of precision measurement
 while aiming to provide a new tool for understanding both astrophysics and fundamental physics.
Squeezed states of light offer a proven means of increasing the sensitivity of \gw\ detectors,
 potentially increasing the rate at which astrophysical sources are detected by more than an order of magnitude.
Since radiation pressure noise plays an important role in advanced detectors,
 frequency dependent squeezing will be required.
In this paper we propose a practical approach to producing frequency dependent squeezing
 for Advanced LIGO and similar interferometric \gw\ detectors.
\end{abstract}

%%%%%%%%%%%%%%%%%%%%
\maketitle

%%%%%%%%%%%%%%%%%%%%%%%%%%%%%%%%%%%%%%%%
%%%%%%%%%%%%%%%%%%%%%%%%%%%%%%%%%%%%%%%%
\section{Introduction}

The Laser Interferometer Gravitational-Wave Observatory (LIGO)
 is part of a global effort to directly detect \gws,
 which has the potential to revolutionize our understanding of both astrophysics,
 and fundamental physics.\cite{httpLIGO, httpVirgo, httpKAGRA, httpGEO}
To realize this potential fully, however,
 significant improvements in sensitivity will be needed beyond Advanced LIGO
 and similar detectors.\cite{Harry2010, GWIC_roadmap}

The use of squeezed states of light (known simply as ``squeezing'') offers a promising direction for sensitivity
 improvement, and has the advantage of requiring minimal changes to the
 detectors currently under construction.\cite{GEO_SQZ_2011, Barsotti2013} 
Squeezing in advanced detectors will require that the squeezed state be
 varied as a function of frequency to suppress both the low-frequency radiation pressure noise and the 
 high-frequency shot noise.\cite{Kimble2001, Chelkowski2005}
When a squeezed state is reflected off a detuned optical cavity, the frequency-dependent amplitude and phase response of the cavity can be used to vary the squeezed state as a function of frequency.
Though all future detectors plan to reduce quantum noise through squeezing,
 uncertainty remains with respect the practical design and limitations
 of the resonant optical cavities, known as ``filter cavities'', this will require.\cite{McClelland2011,Khalili2010}

In this paper we propose a practical approach to producing frequency dependent squeezing
 as required by advanced interferometric \gw\ detectors.
Section \sref{filt_cav} describes the impact of squeezing and filter cavities
 in the context of a detector with realistic thermal noise,
 preparing us for a down-selection among filter cavity topologies.
Section \sref{aLIGO} goes on to suggest a filter cavity implementation
 appropriate for Advanced LIGO in light of realistic optical losses and other practical factors.
Finally, in Appendix \sref{math} we introduce a new approach to computing the
 quantum noise performance of interferometers with filter cavities in the presence
 of multiple sources of optical loss.

%%%%%%%%%%%%%%%%%%%%%%%%%%%%%%%%%%%%%%%%
%%%%%%%%%%%%%%%%%%%%%%%%%%%%%%%%%%%%%%%%
\MEsec{Quantum Noise Filtering in Advanced \GW\ Detectors}{filt_cav}

%%%%%%%%%%
\MEfig{t!}{0.45}{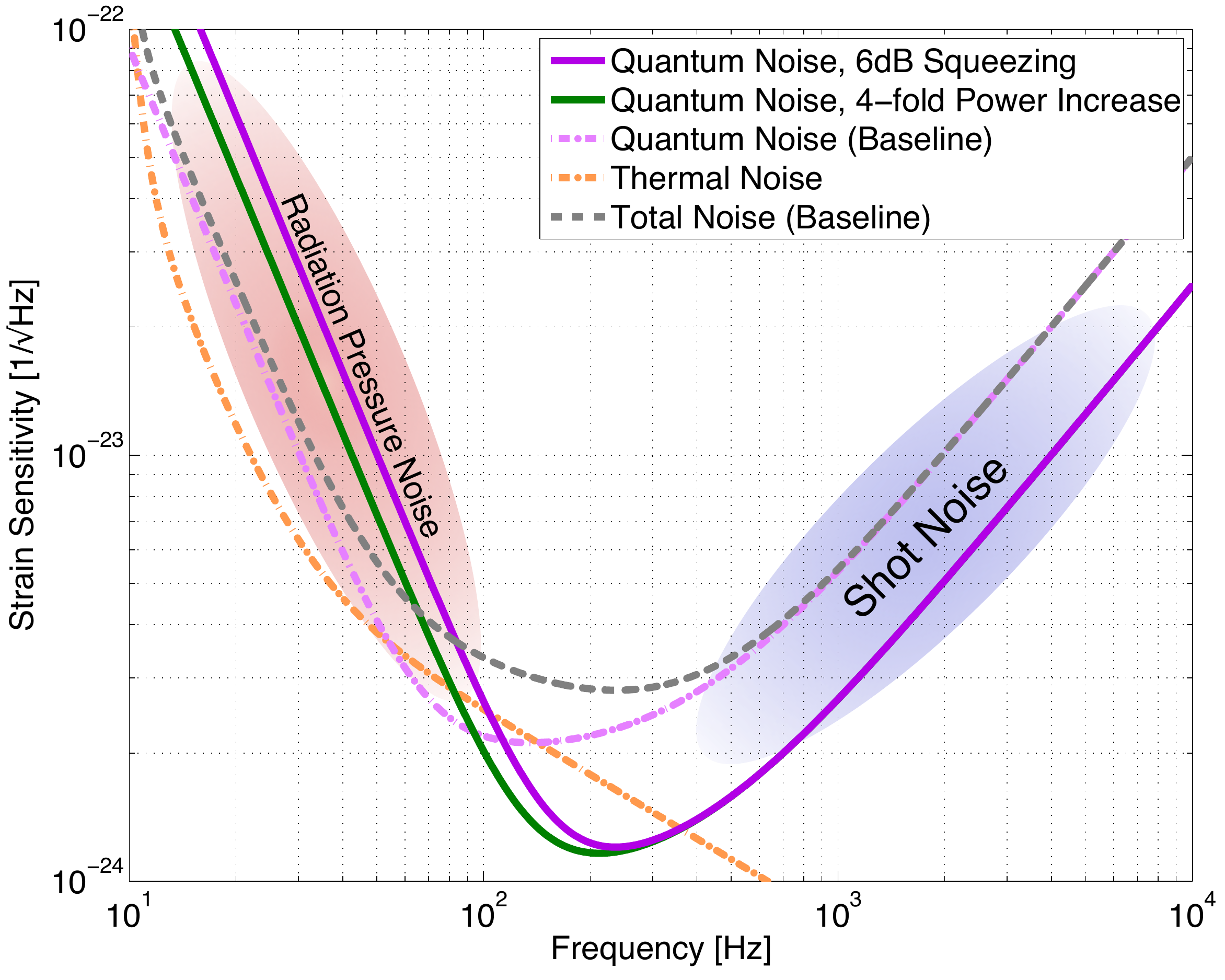}
{The sensitivity of an Advanced LIGO-like detector, our baseline (see table \tref{symbols} for parameters),
 is limited at most frequencies
 by quantum noise, though at low frequencies thermal noise also contributes significantly.
In terms of quantum noise, 
 6 dB of squeezing is equivalent to a 4-fold increase in power circulating in the interferometer,
 though always somewhat worse at low frequencies due to degradation of the squeezed
 vacuum state by optical losses and technical noises.\cite{Dwyer2013}
All of the curves in this paper assume $5\%$ injection loss (e.g., from the squeezed vacuum source to the interferometer),
 and  $10\%$ readout loss (e.g., from the interferometer to the readout,
 including the photodetector quantum efficiency).}
%%%%%%%%%%

The quantum mechanical nature of light sets fundamental constraints on our ability
 to use it as a measurement tool, and in particular it produces a noise floor in interferometric
 position measurements like those employed by \gw\ detectors.
Figure \fref{SQZ_without_FC} shows the expected sensitivity of an Advanced LIGO-like detector,
 in which quantum noise is dominant at essentially all frequencies
 (see Appendix \sref{math} for calculation method and parameters).

The injection of a squeezed vacuum state into an interferometer can reduce quantum noise,
 as demonstrated in GEO600 and LIGO.\cite{GEO_SQZ_2011, Barsotti2013}
The ``frequency independent'' form of squeezing used in these demonstrations will \emph{not},
 however, result in a uniform sensitivity improvement in an advanced detector.
The difference comes from the expectation that radiation pressure noise will play a significant
 role in advanced detectors, and squeezing reduces shot noise at the expense of increased
 radiation pressure noise (see figure \fref{SQZ_without_FC}).

These demonstration experiments focused on reducing quantum noise at high frequency,
 where shot noise dominates, giving results similar to increasing the total power circulating
 in the detector (reduced shot noise and increased radiation pressure noise).
Increasing circulating power, however, leads to significant technical difficulties related
 to thermal lensing and parametric instabilities.\cite{Evans2010, Harry2010}
Furthermore, squeezing offers a means of reducing the quantum noise
 at \emph{any frequency} simply by rotating the squeezed quadrature relative to
  the interferometer signal quadrature (known as the ``squeeze angle'').\cite{Harms2003}
In general, the squeeze angle that minimizes quantum noise varies as a function of frequency,
 and a technique for producing the optimal angle at \emph{all frequencies}
 is required to make effective use of squeezing.
One such technique is to reflect a frequency independent squeezed state off of a detuned Fabry-Perot cavity,
 known as a ``filter cavity''.\cite{Kimble2001}
Figure \fref{IFO_with_FC} shows a simplified interferometer with a squeezed light source and a filter cavity.

%%%%%%%%%%
\MEfig{t!}{0.4}{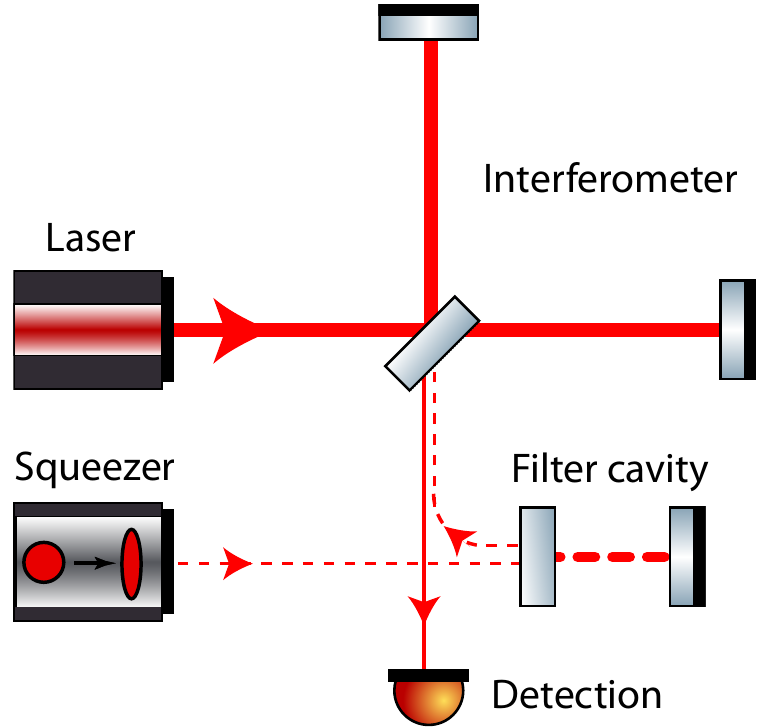}{Frequency dependent squeezing can be produced with the help of a filter cavity.
	Shown here is a simplified interferometer with a linear ``input filter cavity''.\cite{Kimble2001}}
%%%%%%%%%%

There are two primary options for the use of filter cavities to improve the sensitivity of advanced \gw\ detectors;
 rotation of the squeezed vacuum state before it enters the interferometer, known as ``input filtering'',
 and rotation of the signal and vacuum state as they exit the interferometer,
 known as ``variational readout'' or ``output filtering''.\cite{Kimble2001,Harms2003,Khalili2008}
While variational readout appears to have great potential,
 a comprehensive study of input and output filtering in the
 presence of optical losses found that they result in essentially
 indistinguishable detector performance.\cite{Chen2010,McClelland2011,Khalili2010}

%%%%%%%%%%
\MEfig{t!}{0.45}{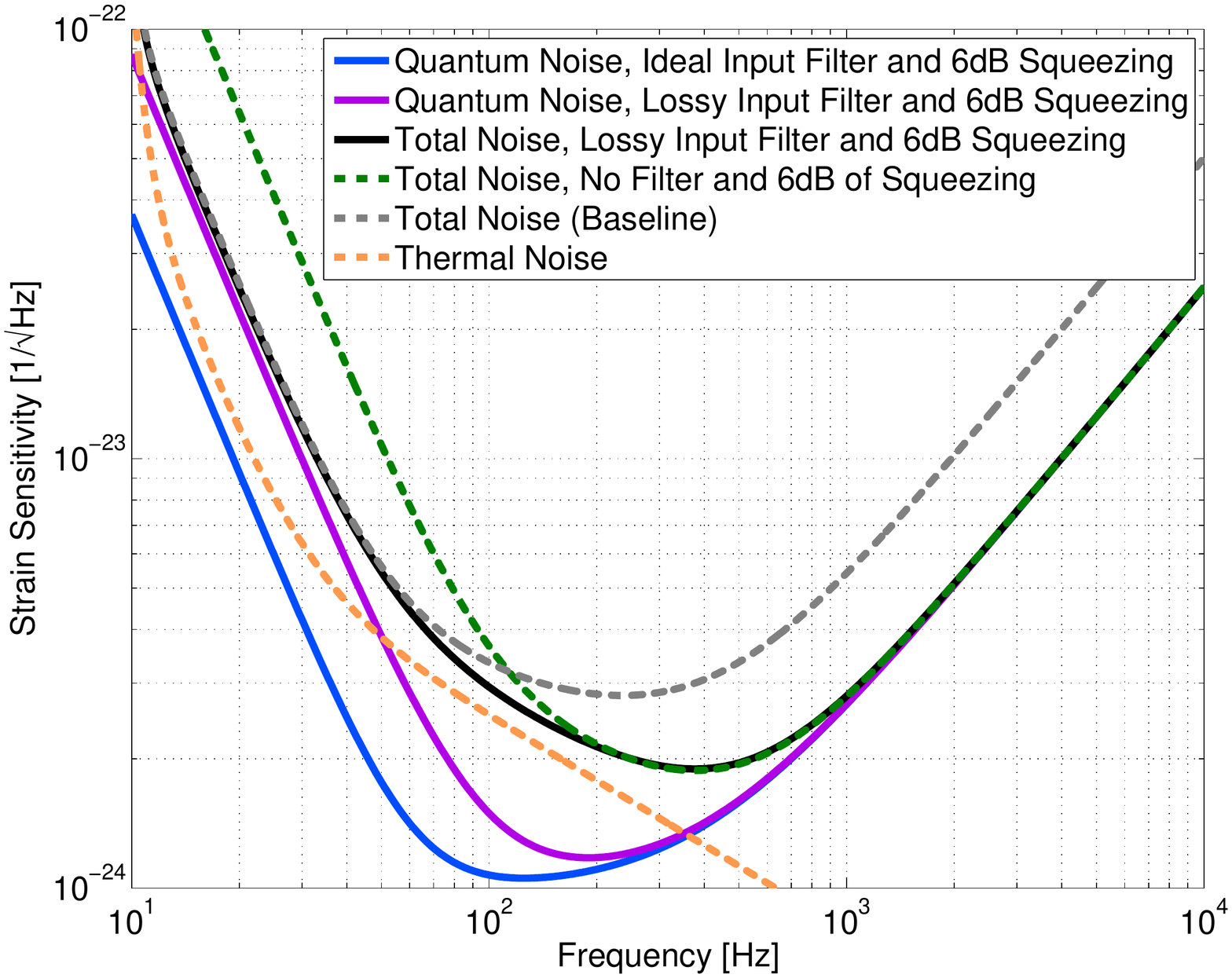}
{Frequency dependent squeezing
 decreases both shot noise and radiation pressure noise,
 reducing quantum noise at all frequencies (blue curve).
A lossy filter cavity has degraded performance in the radiation pressure dominated region
 relative to an ideal lossless filter cavity (purple curve).
The lossy filter shown here, with \SI{1}{ppm/\meter} round-trip loss,
 represents a significant advantage over frequency independent squeezing
 in that it prevents an \emph{increase} in radiation pressure noise
 (green dashed curve, see figure \fref{SQZ_without_FC}).
Furthermore, since thermal noise is significant in the region where radiation pressure noise
 acts, there is little to be gained by making a lower loss filter cavity in the context of a near-term
 upgrade to Advanced LIGO.}
%%%%%%%%%%

%%%%%%%%%%%%%%%%%%%%%%%%%%%%%%%%%%%%%%%%
%%%%%%%%%%%%%%%%%%%%%%%%%%%%%%%%%%%%%%%%
\MEsec{Filtering Solution  for Advanced LIGO}{aLIGO}

In this section we propose a \emph{realistic} filter cavity arrangement for implementation
 as an upgrade to Advanced LIGO.
The elements of realism which drive this proposal are;
 the level of thermal noise and mode of operation expected in Advanced LIGO,
 the geometry of the Advanced LIGO vacuum envelope,
 and achievable values for optical losses in the squeezed field injection chain,
 in the filter cavity, and in the readout chain.

While frequency dependent squeezing for a general signal recycled interferometer requires two filter cavities,
 only one filter cavity is required to obtain virtually
 optimal results for a wide-band interferometer that is operated on resonance
 (i.e. in a ``tuned'' or ``broadband'' configuration).\cite{Purdue2002, Khalili2010}
Since this is likely to be the primary operating state of Advanced LIGO
 we will start by restricting our analysis to this configuration.

%%%%%%%%%%
%\MEfig{t!}{0.45}{SQZ_lossy_FC}{Frequency dependent squeezing, produced with a lossy filter cavity,
%	 has degraded performance in the radiation pressure dominated region
%	 relative to a lossless filter cavity (see figure \fref{SQZ_with_FC}).
%	 A filter cavity with as much as \SI{1}{ppm/\meter} round-trip loss of represents a
%	 significant advantage over frequency independent squeezing
%	 in that it prevents an \emph{increase} in radiation pressure noise
%	 (green dashed curve, see figure \fref{SQZ_without_FC}).
%	 Furthermore, since thermal noise is significant in the region where radiation pressure noise
%	 acts, there is little to be gained by making a lower loss filter cavity in the context of a near-term
%	 upgrade to Advanced LIGO (blue dashed and black curves).\vspace{-8pt}}
%%%%%%%%%%

%%%%%%%%%%
\MEfig{t!}{0.45}{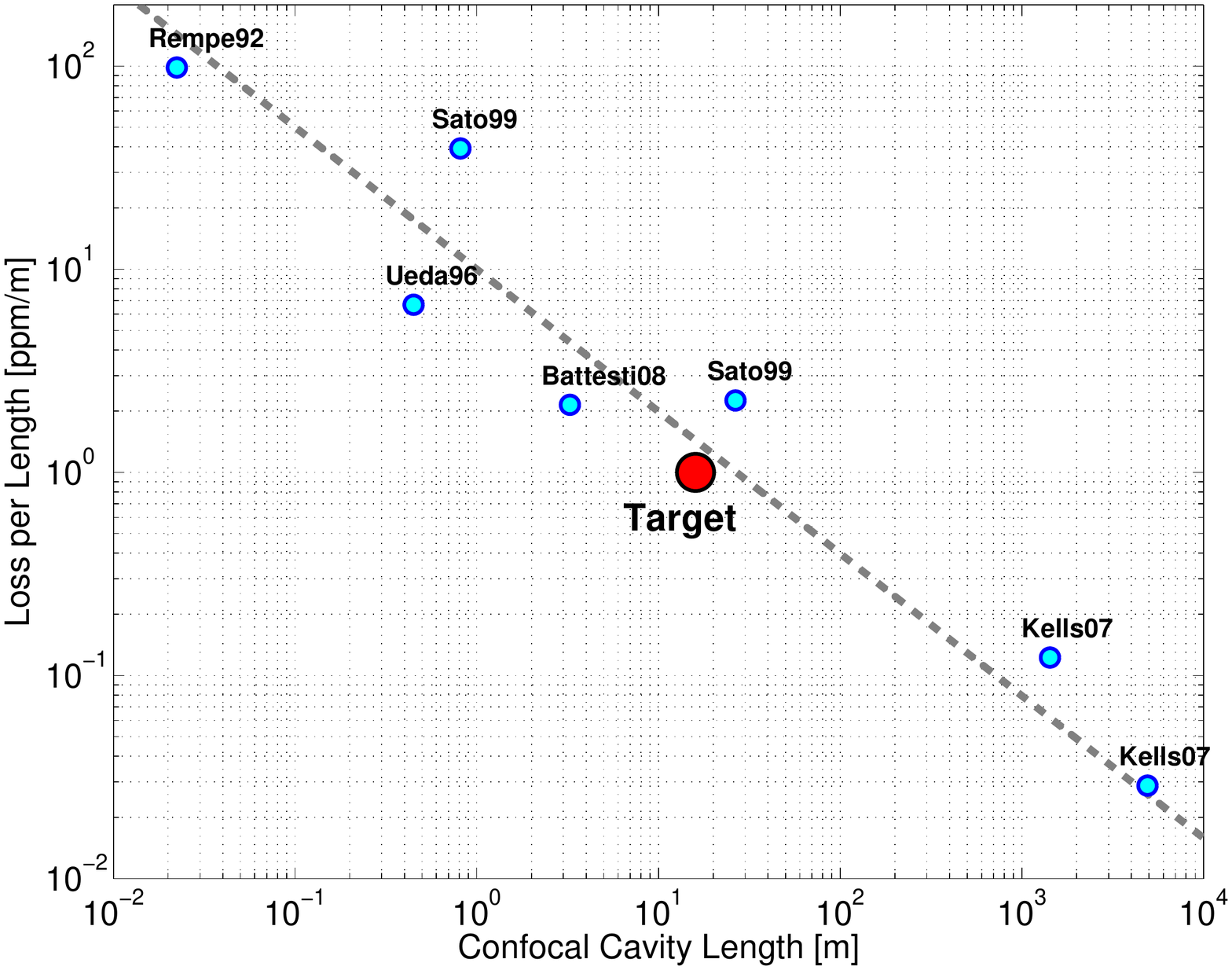}
{Measured round-trip loss per length
 from the literature.\cite{Rempe1992,Ueda1996,Sato1999,Kells2007,Battesti2007}
Losses grow with beam size on the optics,
 so a confocal geometry is optimal for minimizing losses
 and is a good choice for a filter cavity.\cite{Magana2012}
To remove any dependence on the choice of cavity geometry in the experiments
 presented here, the beam sizes on the optics are used to scale the cavity length
 to that of an equivalent confocal cavity.
A rough fit to these data is included to guide the eye, and our target value
of \SI{1}{ppm/\meter} in a \SI{16}{\meter} cavity is marked.}
%%%%%%%%%%

Practically speaking, input filtering has the advantage of being functionally
 separate from the interferometer readout.
That is, input filtering can be added to a functioning \gw\ detector without modification,
 and even after it is added its use is elective.
The same cannot be said for output filtering, which requires that a filter cavity be inserted into the
 interferometer's readout chain.
Furthermore, variational readout is incompatible with current DC readout schemes which produce
 a carrier field for homodyne detection by introducing a slight offset into the differential arm length,
 essentially requiring that the homodyne readout angle match the signal at DC,
 which is orthogonal to the angle required by variational readout.\cite{Kimble2001,Fricke2012,Evans2013}

The essentially identical performance of input and output filtering,
 especially in the presence of realistic thermal noise,
 combined with the practical implications of both schemes push us to
 select a \textbf{single input filter cavity} as a the preferred option.
That said, the requirements which drive input and output filter cavity design
 are very similar such that given a compatible readout scheme the
 input filter cavity solution presented here can also be used as an
 output filter cavity.

The impact of losses on cavity line-width is inversely proportional to cavity length,
 and thus it is always the \emph{loss per unit length} that determines filter cavity performance.\cite{Khalili2010}
Figure \fref{SQZ_with_FC} shows that less than \textbf{1 ppm/m round-trip loss} will not significantly
 improve the detector's sensitivity, so we take this as the target for an Advanced LIGO filter cavity.
While this appears within reach given the optical losses reported in the literature (see figure \fref{loss_vs_length}),
 a study of optical losses in high finesse cavities as a function of cavity length with currently available polishing
 and coating technologies will be required to determine if this is realizable.

To achieve this loss target it is important to choose a filter cavity geometry which minimizes losses.
A \textbf{2-mirror linear cavity} with an isolator is a better choice than the
 3-mirror triangular geometry frequently used to represent filter cavities,\cite{Kimble2001,Khalili2007,Khalili2010}
 since they both have essentially the same length
 while the linear cavity has fewer reflections and thus lower loss
 (see figure \fref{FC_topologies} for details).

Looking to a near-term upgrade of Advanced LIGO,
 the implementation of a short filter cavity which can be housed in the existing
 vacuum envelope will make squeezing a very attractive option.
Since loss per unit length is observed to decrease with cavity length,
 it is advantageous to make a filter cavity as long as possible.
The largest usable distance between two vacuum chambers that house the readout optics
 in the main experimental hall of Advanced LIGO is about \SI{16}{\meter} (see figure \fref{FC_aLIGO}),
 which is plausibly sufficient to achieve the desired 1 ppm/m loss target.
 
Several technical issues in the implementation of a filter cavity for
 Advanced LIGO remain to be addressed before a precise performance prediction can be made.
In addition to measurements of the dependence of optical loss on cavity length,
the impact of spatial mode-mismatch and other sources of loss will need to be carefully evaluated.
An analysis of technical noises which degrade filter cavity performance,
 especially via noise in the reflection phase of the filter cavity, will also be required.

%%%%%%%%%%
\MEfig{t!}{0.47}{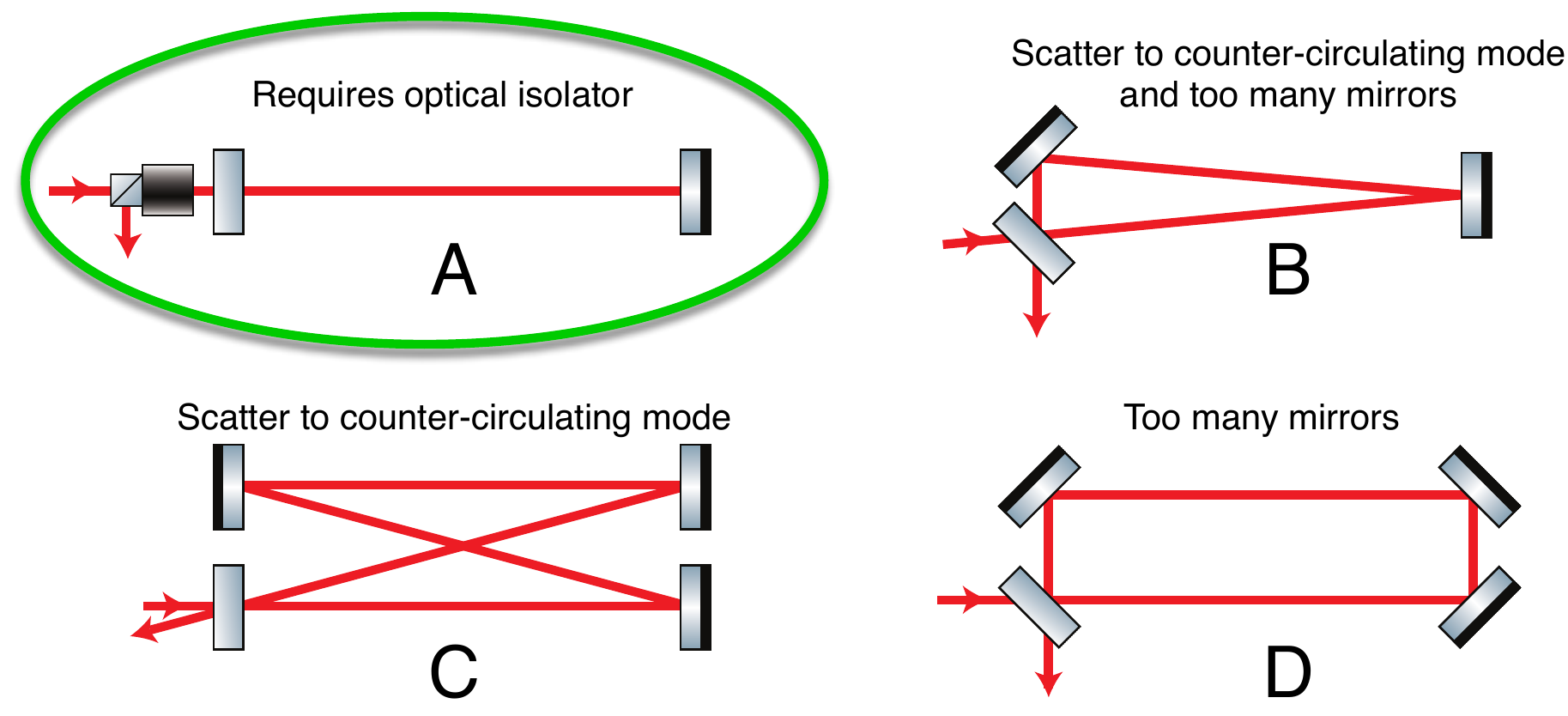}
{Various geometries considered for implementing a filter cavity.
Factors considered in choosing among these geometries were:
 cavity length per mirror, potential for scatter into a counter-circulating mode,
 mode quality degradation due to large angle reflection from curved optics,
 and the ease of separating incident and reflected fields.
While a linear cavity (option A) requires an optical isolator to separate input and output fields,
 it has the best length per mirror, no counter circulating mode, and good mode quality.
A polarization based optical isolator is likely to introduce at least $1\%$ loss in the squeezed field injection.
For short filter cavities, in which the separation between input and output beams
 is significant in a bow-tie geometry (C), the coupling to a counter-circulating
 mode may be negligible and the absence of an optical isolator will make this option preferable.}
%%%%%%%%%%

%%%%%%%%%%%%%%%%%%%%%%%%%%%%%%%%%%%%%%%%
%%%%%%%%%%%%%%%%%%%%%%%%%%%%%%%%%%%%%%%%
\MEsec{Conclusions}{conclusions}

Squeezed vacuum states offer a proven means of enhancing the sensitivity of \gw\ detectors,
 potentially increasing the rate at which astrophysical sources
 are detected by more than an order of magnitude.\cite{GEO_SQZ_2011, Barsotti2013} 
However, due to radiation pressure noise advanced detectors
 pose a more difficult problem than the proof-of-principle experiments conducted to date,
 and narrow line-width optical cavities will be required to make effective use of squeezing.

This work adds to previous efforts,
 which were aimed at parameterizing optimal filter cavity
 configurations in the context of a generalized interferometric \gw\ detector,
 by identifying a practical and effective filter cavity design
 for Advanced LIGO and similar interferometric detectors.
We find that a two-mirror linear cavity with \SI{1}{ppm/\meter} round-trip loss
 is sufficient to reap the majority of the benefit that squeezing can provide.

Finally, we include a mathematical formalism which can be used
 to compute quantum noise in the presence of multiple sources of optical loss.
This approach also maps the ``audio sideband'' or ``one-photon'' formalism
 commonly used to compute the behavior of classical fields in interferometers
 to the two-photon formalism used to compute quantum noise,
 thereby offering a simple connection between the optical parameters
 of a system and its quantum behavior.

%%%%%%%%%%
\MEfig{t!}{0.38}{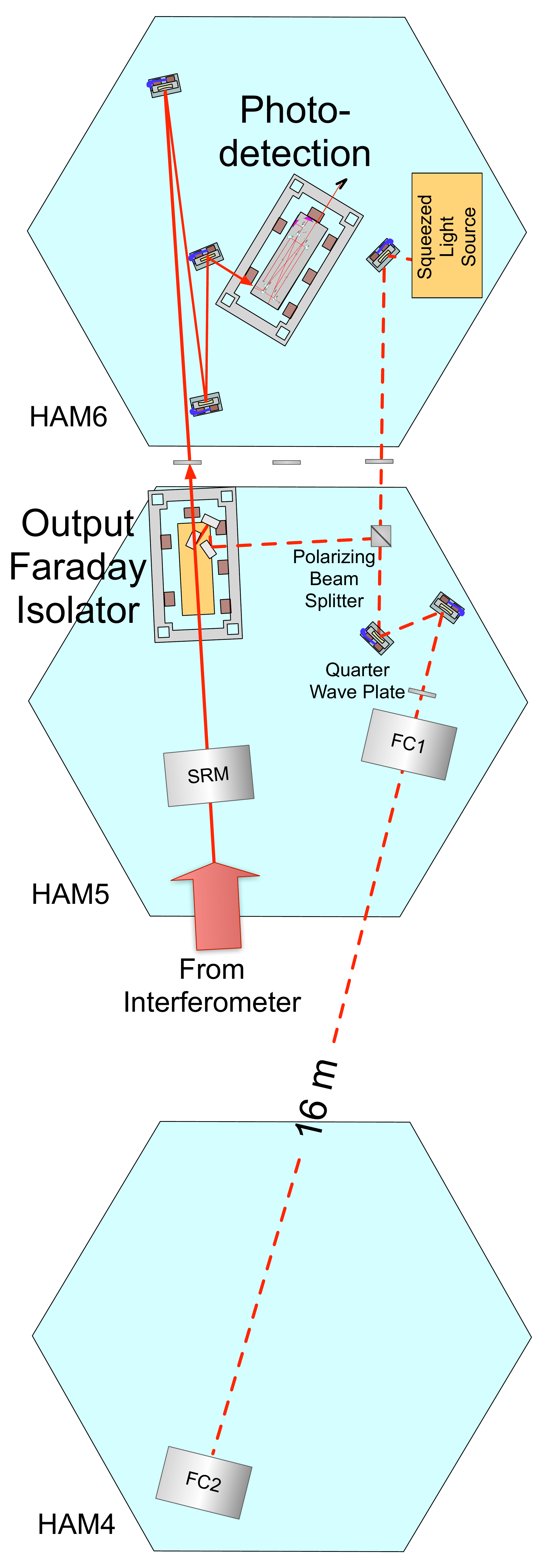}{An input filter cavity could be included
 in the existing vacuum envelope of Advanced LIGO.
The HAM4 and HAM5 chambers shown in this figure contain the interferometer output optics,
 and offer a \SI{16}{\m} baseline with superior vibration isolation and direct access to the Faraday
 isolator with which squeezed vacuum will be injected.
Acronyms used in the figure: signal recycling mirror (SRM), filter cavity input mirror (FC1),
 filter cavity end mirror (FC2).\vspace{-10pt}}
%%%%%%%%%%

%%%%%%%%%%%%%%%%%%%%%%%%%%%%%%%%%%%%%%%%
%%%%%%%%%%%%%%%%%%%%%%%%%%%%%%%%%%%%%%%%
\begin{acknowledgments}
The authors gratefully acknowledge the support of the National Science Foundation and
the LIGO Laboratory, operating under cooperative agreement PHY-0757058.
The authors also acknowledge the wisdom and carefully aimed gibes received
 from Yanbei Chen, Nergis Mavalvala, and Rana Adhikari,
 all of which helped to motivate and mold the contents of this work.
 Jan Harms carried out his research for this paper at the California Institute of
Technology.
This paper has been assigned LIGO Document Number LIGO-P1300054.
\end{acknowledgments}

%%%%%%%%%%%%%%%%%%%%%%%%%%%%%%%%%%%%%%%%
%%%%%%%%%%%%%%%%%%%%%%%%%%%%%%%%%%%%%%%%
\appendix

%%%%%%%%%%%%%%%%%%%%%%%%%%%%%%%%%%%%%%%%
%%%%%%%%%%%%%%%%%%%%%%%%%%%%%%%%%%%%%%%%
\section{Mathematical Formalism}
\sslabel{math}

We use the following mathematical formalism to calculate quantum noise,
 as shown in figures \fref{SQZ_without_FC} and \fref{SQZ_with_FC}.
This treatment is
 based on Kimble et at.,\cite{Kimble2001} Buonanno and Chen,\cite{Buonanno2001}
 and Harms et al.\cite{Harms2003} (hence forth KLMTV, BnC and HCF).
The notation of HCF is used whenever possible.

This treatment extends HCF to a simple means of including all sources of loss,
 but makes no attempt to find analytic expressions for optimal values for filter cavity
 parameter as this appears better done numerically.
To simplify the relevant expressions,
 a transformation from the one-photon formalism to the two-photon formalism is also presented,
 and utilized for computing the input-output relations of filter cavities.\cite{Caves1985}

We start by noting that equation 2 in HCF can be written to express the output of the
 interferometer as a sum of signal and noise fields
 \beq{output_field}
 \Fout = \Fsig ~ h + \sum_n \Tn \Fin
 \eeq
 as suggested in the text preceding their equation 1.
(The pre-factor $1 / M$ in HCF is absorbed into $\Tn$ and $\Fsig$ to make them more independent.) 
In this notation each $\Fin$ is a coherent vacuum field
 of the two-photon formalism such that
\beq{Fin}
\Fin = \smallMatrix{\alpha_1 & \alpha_2}, ~
 \alpha_1 = \smallMatrix{1 \\ 0}, ~
 \alpha_2 = \smallMatrix{0 \\ 1}
\eeq
 where $\alpha_1$ and $\alpha_2$ represent the two field quadratures.\cite{Yuen1976, Caves1985}
Thus, for $N$ vacuum fields entering an interferometer $\set{ \FinN{1}, \dots, \FinN{N}}$,
 there are $2 N$ independent noise sources.
 
Carrying this through to the noise spectral density (HCF equation 7) gives
 \beq{quantum_noise}
 S_h = \frac{\sum_n \abs{\Fhd  \cdot \Tn}^2 }{\abs{\Fhd \cdot \Fsig}^2}
 \eeq
 where we have kept the homodyne field vector $\Fhd$ as in BnC.
To avoid ambiguity: in our notation the dot product of vectors
 $\mathbf{\bar{a}} \cdot \mathbf{\bar{b}} \equiv \left< \mathbf{a} | \mathbf{b} \right> \equiv \sum_n a_n^* b_n$
 is a complex number, the dot product of a vector and a matrix
 $\mathbf{\bar{a}} \cdot \mathbf{B} \equiv \left< \mathbf{a} \right| \mathbf{B} \equiv  \mathbf{\bar{a}}^\dagger \mathbf{B}$
 is a vector, and the magnitude squared of a vector $\abs{\mathbf{\bar{a}}}^2 \equiv \sum_n \abs{a_n}^2$
  is a real number.

In equation \eref{quantum_noise}, each $\Tn$ is a $2 \! \times \! 2$ the transfer matrix which takes
 the coherent vacuum field $\Fin$ from its point of entry into the interferometer to the readout photodetector.
The transfer matrix $\TnN{1}$ for vacuum fluctuations from the squeezed light source, for instance,
 can be constructed by taking a product of transfer matrices
\beq{transfer_SQZa}
\TnN{1} = \TnN{ro}  \; \TnN{ifo} \; \TnN{inj} \; S(r, \lambda)
\eeq
 where $S(r, \lambda)$ is the operator for squeezing by $e^r$ with angle $\lambda$,
 $\TnN{inj}$ takes the squeezed field from its source to the interferometer,
 $\TnN{ifo}$ is the input-output transfer matrix of the interferometer ($\mathbf{C}/M$ in BnC, $\mathbf{T}/M$ in HCF),
 and $\TnN{ro}$ transfers the field from the interferometer to the readout.
In the case of input filtering $\TnN{inj}$ will impose a frequency dependent rotation of the squeeze angle,
 while for output filtering $\TnN{ro}$ will rotate the noise and signal fields as they propagate to the photodetector.
 
 Computation of the input-output relation for any optical system can be performed in the
  two-photon formalism,\cite{Corbitt2005} or in the simpler audio-sideband ``one-photon'' picture
  where the transfer of a field at frequency $\omega_0 + \Omega$ is given
  by $\tau( \Omega )$  (as in previous works $\omega_0$ is the carrier frequency and
  $\Omega$ the signal sideband frequency). 
To convert between pictures we define transfer coefficients for positive and negative sidebands
 as $\tau_+ = \tau( \Omega )$ and $\tau_- = \tau( -\Omega )$, and then compute the
 two-photon transfer matrix as
\beq{one_photon}
  \mathbf{T} = \TwoP \smallMatrix{\tau_+ & \\ & \tau^*_-} \TwoP^{-1},  ~
  \TwoP = \frac{1}{\sqrt{2}} \smallMatrix{1 & 1 \\ -i & +i}~.
%  \TwoP^{-1} = \frac{1}{\sqrt{2}} \smallMatrix{1 & +i \\ 1 & -i}.
\eeq
Off-diagonal elements in the one-photon transfer matrix are unnecessary for filter cavities,
 since those elements represent non-linear transformations
 that couple the positive and negative sideband amplitudes
 such as squeezing or radiation pressure back-action.

For a filter cavity, the one-photon transfer coefficient is just the amplitude reflectivity of the cavity
\beq{fc_refl}
r_{fc}( \Omega ) = r_{in} - \frac{t_{in}^2}{r_{in}} \; \frac{r_{rt} e^{-i \phi(\Omega)}}{1 - r_{rt} e^{-i \phi(\Omega)}}
\eeq
 where the cavity round-trip reflectivity $r_{rt}$ is related to its bandwidth by
\beq{fc_bandwidth}
\gamma_{fc} = -\log(r_{rt}) \; f_{FSR} \simeq \frac{1 - r_{rt}^2}{2} f_{FSR}, \quad f_{FSR} = \frac{c}{2 L_{fc}}
\eeq
 and the round-trip phase derives from the cavity detuning according to
\beq{fc_detuning}
\phi(\Omega) = \frac{\Omega - \Delta \omega_{fc}}{f_{FSR}}, \quad  \Delta \omega_{fc}= \omega_{fc} - \omega_0
\eeq
 for a cavity of length $L_{fc}$ and resonant frequency $\omega_{fc}$.
 (The cavity ``bandwidth'' is the half-width of the resonance, a.k.a. the ``cavity pole'' frequency.
 In KLMTV the detuning is given in terms of the bandwidth $\xi = \Delta \omega_{fc} / \gamma_{fc}$.)
The amplitude reflectivity of the cavity input-output coupler is related to its transmissivity by
$r_{in}^2 \leq 1 - t_{in}^2$, and to the round-trip reflectivity by $r_{rt} \leq r_{in}$,
 where a lossless filter cavity attains equality in both expressions.
The filter cavity transfer matrix in the two-photon formalism is
\beq{audio_transfer}
  \TnN{fc} = \TwoP \smallMatrix{r_{fc+} & \\ & r^*_{fc-}} \TwoP^{-1}.
\eeq

Returning to an interferometer with input filtering,
 and including injection and readout losses, we can write 
\beqa{transfer_SQZb}
\Fsig & = & L(\Lambda_{ro})  \; \Fsifo \\
\TnN{in 1} & = & L(\Lambda_{ro})  \; \Tifo \; \Tfc \; L(\Lambda_{inj}) \; S(r, \lambda) \\
\TnN{in 2} & = & L(\Lambda_{ro})  \; \Tifo \; \Tfc \; \Lambda_{inj} \\
\TnN{in 3} & = & L(\Lambda_{ro})  \; \Tifo\; \Lambda_{fc} \\
\TnN{in 4} & = & \Lambda_{ro}
\eeqa
 where $L(\Lambda_x) = \sqrt{1 - \Lambda_x^2}$ is the transfer coefficient of a power loss $\Lambda_x^2$.
The four $\Tn$ terms each contribute to the sum in equation \eref{quantum_noise},
 since each represents an entry point for coherent vacuum fluctuations.\cite{Kimble2001}
$\TnN{in 1}$ propagates the squeezed field from its source to the photodetector,
 and $\TnN{in 2}$ accounts for the losses of the injection path (from squeezer to filter cavity).
$\TnN{in 3}$ represents the frequency dependent losses in the filter cavity with
 \beq{fc_loss}
 \Lambda_{fc}^2 = 1 - (\abs{r_{fc+}}^2 + \abs{r_{fc-}}^2) / 2 \quad .
 \eeq
Finally, $\TnN{in 4}$ accounts for losses in the readout path, from interferometer to photodetector
 including the detector quantum efficiency.

For an interferometer with output filtering, on the other hand, we have
\beqa{transfer_SQZc}
\Fsig & = & L(\Lambda_{ro}) \; \Tfc  \; \Fsifo \\
\TnN{out 1} & = & L(\Lambda_{ro}) \; \Tfc \; \Tifo \; L(\Lambda_{inj}) \; S(r, \lambda) \\
\TnN{out 2} & = & L(\Lambda_{ro}) \; \Tfc \; \Tifo \; \Lambda_{inj} \\
\TnN{out 3} & = & L(\Lambda_{ro})  \; \Lambda_{fc} \\
\TnN{out 4} & = & \Lambda_{ro}
\eeqa
 where the signal produced by the interferometer $\Fsifo$ is modified by
 the filter cavity, in addition to experiencing some loss in the readout process.

The transfer matrix and signal vector for a tuned signal recycled interferometer
 of the sort discussed in section \sref{aLIGO} and analyzed in III.C.2 of BnC, are
\beq{Tifo}
\Tifo = \smallMatrix{1 & \Ksr \\ 0 & 1}, ~
\Fsifo = \frac{\sqrt{2 \Ksr}}{h_{SQL}}\smallMatrix{1 \\ 0}, ~
\eeq
 where
 \begin{gather}
 \elabel{ifo_eqs}
h_{SQL} = \sqrt{\frac{8 \hbar}{m L_{arm}^2 \Omega^2}}, \quad
\Ksr = \frac{\Kifo \; t_{sr}^2}{\abs{1 + e^{2 i \Phi} r_{sr}}^2}, \\
\Kifo = \frac{8 P_{bs} \; \omega_0}{m L_{arm}^2 \Omega^2 (\Omega^2 + \gamma_{arm}^2)}, \\
\Phi = \Omega L_{src} / c + \tfun{arctan}{\Omega / \gamma_{arm}},
\end{gather}
 and the signal recycling mirror amplitude reflectivity and transmissivity
 are $r_{sr}$ and $t_{sr}$ for consistency with equations \eref{fc_refl}-\eref{audio_transfer}
 (these are $\rho$ and $\tau$ in BnC and HCF). (There appear to be several errors in HCF with
  respect to the equations for Advanced LIGO, making BnC a better reference.)
Symbol definitions, and the values used in our calculations, are given in table \tref{symbols}.

\begin{table}
\caption{\tlabel{symbols}Symbols and values}
\begin{ruledtabular}
\begin{tabular}{lcr}
Symbol & Meaning & Value \\
\hline
$c$ & light speed & \SI{299792458}{\m/\s} \\
$\omega_0$ & frequency of carrier field & $2 \pi \times \SI{282}{\THz}$\\
$P_{bs}$ & power on the beam-splitter & \SI{5.6}{\kW} \\
$m$ & mass of each test-mass mirror & \SI{40}{\kg} \\
$L_{arm}$ & arm cavity length & \SI{3995}{\meter} \\
$L_{src}$ &  signal cavity length & \SI{55}{\meter} \\
$\gamma_{arm}$ & arm cavity half-width & $2 \pi \times \SI{42}{\Hz}$\\
$t_{sr}^2$ & signal mirror power transmission & $35\%$  \\
$\Lambda_{inj}^2$ & injection losses & $5\%$ \\
$\Lambda_{ro}^2$ & readout losses & $10\%$ \\
$t_{in}^2$ & filter cavity input transmission & \\
 & ideal \SI{16}{\m} filter cavity & \SI{66}{ppm}  \\
 & \SI{1}{ppm/\meter} loss \SI{16}{\m} filter cavity & \SI{50}{ppm}  \\
$\gamma_{fc}$ & filter cavity half-width & $2 \pi \times \SI{49}{\Hz}$\\
$\Delta \omega_{fc}$ & filter cavity detuning & $2 \pi \times \SI{49}{\Hz}$
\end{tabular}
\end{ruledtabular}
\end{table}

%%%%%%%%%%%%%%%%%%%%%%%%%%%%%%%%%%%%%%%%
% BIB
%%%%%%%%%%%%%%%%%%%%%%%%%%%%%%%%%%%%%%%%
%\newpage
%\setcounter{page}{1}
%\bibliographystyle{plain}

\bibliography{/Users/mevans/Documents/Work/Latex/papers}

\begin{thebibliography}{31}%
\makeatletter
\providecommand \@ifxundefined [1]{%
 \@ifx{#1\undefined}
}%
\providecommand \@ifnum [1]{%
 \ifnum #1\expandafter \@firstoftwo
 \else \expandafter \@secondoftwo
 \fi
}%
\providecommand \@ifx [1]{%
 \ifx #1\expandafter \@firstoftwo
 \else \expandafter \@secondoftwo
 \fi
}%
\providecommand \natexlab [1]{#1}%
\providecommand \enquote  [1]{``#1''}%
\providecommand \bibnamefont  [1]{#1}%
\providecommand \bibfnamefont [1]{#1}%
\providecommand \citenamefont [1]{#1}%
\providecommand \href@noop [0]{\@secondoftwo}%
\providecommand \href [0]{\begingroup \@sanitize@url \@href}%
\providecommand \@href[1]{\@@startlink{#1}\@@href}%
\providecommand \@@href[1]{\endgroup#1\@@endlink}%
\providecommand \@sanitize@url [0]{\catcode `\\12\catcode `\$12\catcode
  `\&12\catcode `\#12\catcode `\^12\catcode `\_12\catcode `\%12\relax}%
\providecommand \@@startlink[1]{}%
\providecommand \@@endlink[0]{}%
\providecommand \url  [0]{\begingroup\@sanitize@url \@url }%
\providecommand \@url [1]{\endgroup\@href {#1}{\urlprefix }}%
\providecommand \urlprefix  [0]{URL }%
\providecommand \Eprint [0]{\href }%
\providecommand \doibase [0]{http://dx.doi.org/}%
\providecommand \selectlanguage [0]{\@gobble}%
\providecommand \bibinfo  [0]{\@secondoftwo}%
\providecommand \bibfield  [0]{\@secondoftwo}%
\providecommand \translation [1]{[#1]}%
\providecommand \BibitemOpen [0]{}%
\providecommand \bibitemStop [0]{}%
\providecommand \bibitemNoStop [0]{.\EOS\space}%
\providecommand \EOS [0]{\spacefactor3000\relax}%
\providecommand \BibitemShut  [1]{\csname bibitem#1\endcsname}%
\let\auto@bib@innerbib\@empty
%</preamble>
\bibitem [{\citenamefont {{LIGO Laboratory}}(ment)}]{httpLIGO}%
  \BibitemOpen
  \bibfield  {author} {\bibinfo {author} {\bibnamefont {{LIGO Laboratory}}},\
  }\href {http://www.ligo.caltech.edu} {\enquote {\bibinfo {title} {{LIGO} web
  site},}\ } (\bibinfo {year} {living document})\BibitemShut {NoStop}%
\bibitem [{\citenamefont {{Virgo Collaboration}}(ment)}]{httpVirgo}%
  \BibitemOpen
  \bibfield  {author} {\bibinfo {author} {\bibnamefont {{Virgo
  Collaboration}}},\ }\href {http://www.virgo.infn.it} {\enquote {\bibinfo
  {title} {Virgo web site},}\ } (\bibinfo {year} {living document})\BibitemShut
  {NoStop}%
\bibitem [{\citenamefont {KAGRA}(ment)}]{httpKAGRA}%
  \BibitemOpen
  \bibfield  {author} {\bibinfo {author} {\bibnamefont {KAGRA}},\ }\href
  {http://gwcenter.icrr.u-tokyo.ac.jp} {\enquote {\bibinfo {title} {{KAGRA} web
  site},}\ } (\bibinfo {year} {living document})\BibitemShut {NoStop}%
\bibitem [{\citenamefont {GEO600}(ment)}]{httpGEO}%
  \BibitemOpen
  \bibfield  {author} {\bibinfo {author} {\bibnamefont {GEO600}},\ }\href
  {http://www.geo600.de} {\enquote {\bibinfo {title} {{GEO} web site},}\ }
  (\bibinfo {year} {living document})\BibitemShut {NoStop}%
\bibitem [{\citenamefont {Harry}\ and\ \citenamefont {{(The LIGO Scientific
  Collaboration)}}(2010)}]{Harry2010}%
  \BibitemOpen
  \bibfield  {author} {\bibinfo {author} {\bibfnamefont {G.~M.}\ \bibnamefont
  {Harry}}\ and\ \bibinfo {author} {\bibnamefont {{(The LIGO Scientific
  Collaboration)}}},\ }\bibfield  {title} {\enquote {\bibinfo {title}
  {{Advanced LIGO: the next generation of gravitational wave detectors}},}\
  }\href@noop {} {\bibfield  {journal} {\bibinfo  {journal} {Classical and
  Quantum Gravity}\ }\textbf {\bibinfo {volume} {27}},\ \bibinfo {pages}
  {084006} (\bibinfo {year} {2010})}\BibitemShut {NoStop}%
\bibitem [{\citenamefont {{LIGO Scientific
  Collaboration}}(2010)}]{GWIC_roadmap}%
  \BibitemOpen
  \bibfield  {author} {\bibinfo {author} {\bibnamefont {{LIGO Scientific
  Collaboration}}},\ }\href {https://gwic.ligo.org/roadmap/} {\enquote
  {\bibinfo {title} {{GWIC Roadmap}},}\ } (\bibinfo {year} {2010})\BibitemShut
  {NoStop}%
\bibitem [{\citenamefont {{LIGO Scientific
  Collaboration}}(2011)}]{GEO_SQZ_2011}%
  \BibitemOpen
  \bibfield  {author} {\bibinfo {author} {\bibnamefont {{LIGO Scientific
  Collaboration}}},\ }\bibfield  {title} {\enquote {\bibinfo {title} {{A
  gravitational wave observatory operating beyond the quantum shot-noise
  limit}},}\ }\href@noop {} {\bibfield  {journal} {\bibinfo  {journal} {Nature
  Physics}\ }\textbf {\bibinfo {volume} {7}},\ \bibinfo {pages} {962--965}
  (\bibinfo {year} {2011})}\BibitemShut {NoStop}%
\bibitem [{\citenamefont {Barsotti}(2013)}]{Barsotti2013}%
  \BibitemOpen
  \bibfield  {author} {\bibinfo {author} {\bibfnamefont {L.}~\bibnamefont
  {Barsotti}},\ }\href@noop {} {\enquote {\bibinfo {title} {{Enhancing the
  astrophysical reach of the LIGO gravitational wave detector by using squeezed
  states of light}},}\ } (\bibinfo {year} {2013}),\ \bibinfo {note} {in
  Preparation}\BibitemShut {NoStop}%
\bibitem [{\citenamefont {Kimble}\ \emph {et~al.}(2001)\citenamefont {Kimble},
  \citenamefont {Levin}, \citenamefont {Matsko}, \citenamefont {Thorne},\ and\
  \citenamefont {Vyatchanin}}]{Kimble2001}%
  \BibitemOpen
  \bibfield  {author} {\bibinfo {author} {\bibfnamefont {H.}~\bibnamefont
  {Kimble}}, \bibinfo {author} {\bibfnamefont {Y.}~\bibnamefont {Levin}},
  \bibinfo {author} {\bibfnamefont {A.}~\bibnamefont {Matsko}}, \bibinfo
  {author} {\bibfnamefont {K.}~\bibnamefont {Thorne}}, \ and\ \bibinfo {author}
  {\bibfnamefont {S.}~\bibnamefont {Vyatchanin}},\ }\bibfield  {title}
  {\enquote {\bibinfo {title} {{Conversion of conventional gravitational-wave
  interferometers into quantum nondemolition interferometers by modifying their
  input and/or output optics}},}\ }\href@noop {} {\bibfield  {journal}
  {\bibinfo  {journal} {Physical Review D}\ }\textbf {\bibinfo {volume} {65}}
  (\bibinfo {year} {2001})}\BibitemShut {NoStop}%
\bibitem [{\citenamefont {Chelkowski}\ \emph {et~al.}(2005)\citenamefont
  {Chelkowski}, \citenamefont {Vahlbruch}, \citenamefont {Hage}, \citenamefont
  {Franzen}, \citenamefont {Lastzka}, \citenamefont {Danzmann},\ and\
  \citenamefont {Schnabel}}]{Chelkowski2005}%
  \BibitemOpen
  \bibfield  {author} {\bibinfo {author} {\bibfnamefont {S.}~\bibnamefont
  {Chelkowski}}, \bibinfo {author} {\bibfnamefont {H.}~\bibnamefont
  {Vahlbruch}}, \bibinfo {author} {\bibfnamefont {B.}~\bibnamefont {Hage}},
  \bibinfo {author} {\bibfnamefont {A.}~\bibnamefont {Franzen}}, \bibinfo
  {author} {\bibfnamefont {N.}~\bibnamefont {Lastzka}}, \bibinfo {author}
  {\bibfnamefont {K.}~\bibnamefont {Danzmann}}, \ and\ \bibinfo {author}
  {\bibfnamefont {R.}~\bibnamefont {Schnabel}},\ }\bibfield  {title} {\enquote
  {\bibinfo {title} {{Experimental characterization of frequency-dependent
  squeezed light}},}\ }\href@noop {} {\bibfield  {journal} {\bibinfo  {journal}
  {Physical Review A}\ }\textbf {\bibinfo {volume} {71}} (\bibinfo {year}
  {2005})}\BibitemShut {NoStop}%
\bibitem [{\citenamefont {McClelland}\ \emph {et~al.}(2011)\citenamefont
  {McClelland}, \citenamefont {Mavalvala}, \citenamefont {Chen},\ and\
  \citenamefont {Schnabel}}]{McClelland2011}%
  \BibitemOpen
  \bibfield  {author} {\bibinfo {author} {\bibfnamefont {D.}~\bibnamefont
  {McClelland}}, \bibinfo {author} {\bibfnamefont {N.}~\bibnamefont
  {Mavalvala}}, \bibinfo {author} {\bibfnamefont {Y.}~\bibnamefont {Chen}}, \
  and\ \bibinfo {author} {\bibfnamefont {R.}~\bibnamefont {Schnabel}},\
  }\bibfield  {title} {\enquote {\bibinfo {title} {{Advanced interferometry,
  quantum optics and optomechanics in gravitational wave detectors}},}\
  }\href@noop {} {\bibfield  {journal} {\bibinfo  {journal} {Laser {\&}
  Photonics Reviews}\ ,\ \bibinfo {pages} {677--696}} (\bibinfo {year}
  {2011})}\BibitemShut {NoStop}%
\bibitem [{\citenamefont {Khalili}(2010)}]{Khalili2010}%
  \BibitemOpen
  \bibfield  {author} {\bibinfo {author} {\bibfnamefont {F.~Y.}\ \bibnamefont
  {Khalili}},\ }\bibfield  {title} {\enquote {\bibinfo {title} {{Optimal
  configurations of filter cavity in future gravitational-wave detectors}},}\
  }\href@noop {} {\bibfield  {journal} {\bibinfo  {journal} {Physical Review
  D}\ }\textbf {\bibinfo {volume} {81}},\ \bibinfo {pages} {122002} (\bibinfo
  {year} {2010})}\BibitemShut {NoStop}%
\bibitem [{\citenamefont {Evans}, \citenamefont {Barsotti},\ and\ \citenamefont
  {Fritschel}(2010)}]{Evans2010}%
  \BibitemOpen
  \bibfield  {author} {\bibinfo {author} {\bibfnamefont {M.}~\bibnamefont
  {Evans}}, \bibinfo {author} {\bibfnamefont {L.}~\bibnamefont {Barsotti}}, \
  and\ \bibinfo {author} {\bibfnamefont {P.}~\bibnamefont {Fritschel}},\
  }\bibfield  {title} {\enquote {\bibinfo {title} {{A general approach to
  optomechanical parametric instabilities}},}\ }\href@noop {} {\bibfield
  {journal} {\bibinfo  {journal} {Physics Letters A}\ }\textbf {\bibinfo
  {volume} {374}},\ \bibinfo {pages} {665--671} (\bibinfo {year}
  {2010})}\BibitemShut {NoStop}%
\bibitem [{\citenamefont {Dwyer}(2013)}]{Dwyer2013}%
  \BibitemOpen
  \bibfield  {author} {\bibinfo {author} {\bibfnamefont {S.}~\bibnamefont
  {Dwyer}},\ }\href@noop {} {\enquote {\bibinfo {title} {{Squeezing angle
  fluctuations in a quantum enhanced gravitational wave detector}},}\ }
  (\bibinfo {year} {2013}),\ \bibinfo {note} {in Preparation}\BibitemShut
  {NoStop}%
\bibitem [{\citenamefont {Harms}\ \emph {et~al.}(2003)\citenamefont {Harms},
  \citenamefont {Chen}, \citenamefont {Chelkowski}, \citenamefont {Franzen},
  \citenamefont {Vahlbruch}, \citenamefont {Danzmann},\ and\ \citenamefont
  {Schnabel}}]{Harms2003}%
  \BibitemOpen
  \bibfield  {author} {\bibinfo {author} {\bibfnamefont {J.}~\bibnamefont
  {Harms}}, \bibinfo {author} {\bibfnamefont {Y.}~\bibnamefont {Chen}},
  \bibinfo {author} {\bibfnamefont {S.}~\bibnamefont {Chelkowski}}, \bibinfo
  {author} {\bibfnamefont {A.}~\bibnamefont {Franzen}}, \bibinfo {author}
  {\bibfnamefont {H.}~\bibnamefont {Vahlbruch}}, \bibinfo {author}
  {\bibfnamefont {K.}~\bibnamefont {Danzmann}}, \ and\ \bibinfo {author}
  {\bibfnamefont {R.}~\bibnamefont {Schnabel}},\ }\bibfield  {title} {\enquote
  {\bibinfo {title} {{Squeezed-input, optical-spring, signal-recycled
  gravitational-wave detectors}},}\ }\href@noop {} {\bibfield  {journal}
  {\bibinfo  {journal} {Physical Review D}\ }\textbf {\bibinfo {volume} {68}},\
  \bibinfo {pages} {042001} (\bibinfo {year} {2003})}\BibitemShut {NoStop}%
\bibitem [{\citenamefont {Khalili}(2008)}]{Khalili2008}%
  \BibitemOpen
  \bibfield  {author} {\bibinfo {author} {\bibfnamefont {F.}~\bibnamefont
  {Khalili}},\ }\bibfield  {title} {\enquote {\bibinfo {title} {{Increasing
  future gravitational-wave detectors' sensitivity by means of amplitude filter
  cavities and quantum entanglement}},}\ }\href@noop {} {\bibfield  {journal}
  {\bibinfo  {journal} {Physical Review D}\ }\textbf {\bibinfo {volume} {77}},\
  \bibinfo {pages} {062003} (\bibinfo {year} {2008})}\BibitemShut {NoStop}%
\bibitem [{\citenamefont {Chen}\ \emph {et~al.}(2010)\citenamefont {Chen},
  \citenamefont {Danilishin}, \citenamefont {Khalili},\ and\ \citenamefont
  {M{\"u}ller-Ebhardt}}]{Chen2010}%
  \BibitemOpen
  \bibfield  {author} {\bibinfo {author} {\bibfnamefont {Y.}~\bibnamefont
  {Chen}}, \bibinfo {author} {\bibfnamefont {S.~L.}\ \bibnamefont
  {Danilishin}}, \bibinfo {author} {\bibfnamefont {F.~Y.}\ \bibnamefont
  {Khalili}}, \ and\ \bibinfo {author} {\bibfnamefont {H.}~\bibnamefont
  {M{\"u}ller-Ebhardt}},\ }\bibfield  {title} {\enquote {\bibinfo {title} {{QND
  measurements for future gravitational-wave detectors}},}\ }\href@noop {}
  {\bibfield  {journal} {\bibinfo  {journal} {General Relativity and
  Gravitation}\ }\textbf {\bibinfo {volume} {43}},\ \bibinfo {pages} {671--694}
  (\bibinfo {year} {2010})}\BibitemShut {NoStop}%
\bibitem [{\citenamefont {Purdue}\ and\ \citenamefont
  {Chen}(2002)}]{Purdue2002}%
  \BibitemOpen
  \bibfield  {author} {\bibinfo {author} {\bibfnamefont {P.}~\bibnamefont
  {Purdue}}\ and\ \bibinfo {author} {\bibfnamefont {Y.}~\bibnamefont {Chen}},\
  }\bibfield  {title} {\enquote {\bibinfo {title} {{Practical speed meter
  designs for quantum nondemolition gravitational-wave interferometers}},}\
  }\href@noop {} {\bibfield  {journal} {\bibinfo  {journal} {Physical Review
  D}\ }\textbf {\bibinfo {volume} {66}} (\bibinfo {year} {2002})}\BibitemShut
  {NoStop}%
\bibitem [{\citenamefont {Fricke}\ \emph {et~al.}(2012)\citenamefont {Fricke},
  \citenamefont {Smith-Lefebvre}, \citenamefont {Abbott}, \citenamefont
  {Adhikari}, \citenamefont {Dooley}, \citenamefont {Evans}, \citenamefont
  {Fritschel}, \citenamefont {Frolov}, \citenamefont {Kawabe}, \citenamefont
  {Kissel}, \citenamefont {Slagmolen},\ and\ \citenamefont
  {Waldman}}]{Fricke2012}%
  \BibitemOpen
  \bibfield  {author} {\bibinfo {author} {\bibfnamefont {T.~T.}\ \bibnamefont
  {Fricke}}, \bibinfo {author} {\bibfnamefont {N.~D.}\ \bibnamefont
  {Smith-Lefebvre}}, \bibinfo {author} {\bibfnamefont {R.}~\bibnamefont
  {Abbott}}, \bibinfo {author} {\bibfnamefont {R.}~\bibnamefont {Adhikari}},
  \bibinfo {author} {\bibfnamefont {K.~L.}\ \bibnamefont {Dooley}}, \bibinfo
  {author} {\bibfnamefont {M.}~\bibnamefont {Evans}}, \bibinfo {author}
  {\bibfnamefont {P.}~\bibnamefont {Fritschel}}, \bibinfo {author}
  {\bibfnamefont {V.~V.}\ \bibnamefont {Frolov}}, \bibinfo {author}
  {\bibfnamefont {K.}~\bibnamefont {Kawabe}}, \bibinfo {author} {\bibfnamefont
  {J.~S.}\ \bibnamefont {Kissel}}, \bibinfo {author} {\bibfnamefont {B.~J.~J.}\
  \bibnamefont {Slagmolen}}, \ and\ \bibinfo {author} {\bibfnamefont {S.~J.}\
  \bibnamefont {Waldman}},\ }\bibfield  {title} {\enquote {\bibinfo {title}
  {{DC readout experiment in Enhanced LIGO}},}\ }\href {\doibase
  10.1088/0264-9381/29/6/065005} {\bibfield  {journal} {\bibinfo  {journal}
  {Classical and Quantum Gravity}\ }\textbf {\bibinfo {volume} {29}},\ \bibinfo
  {pages} {065005} (\bibinfo {year} {2012})}\BibitemShut {NoStop}%
\bibitem [{\citenamefont {Evans}(2013)}]{Evans2013}%
  \BibitemOpen
  \bibfield  {author} {\bibinfo {author} {\bibfnamefont {M.}~\bibnamefont
  {Evans}},\ }\href@noop {} {\enquote {\bibinfo {title} {{Low-Noise Homodyne
  Readout for Quantum Limited Gravitational Wave Detectors}},}\ } (\bibinfo
  {year} {2013}),\ \bibinfo {note} {in Preparation}\BibitemShut {NoStop}%
\bibitem [{\citenamefont {Khalili}(2007)}]{Khalili2007}%
  \BibitemOpen
  \bibfield  {author} {\bibinfo {author} {\bibfnamefont {F.}~\bibnamefont
  {Khalili}},\ }\bibfield  {title} {\enquote {\bibinfo {title} {{Quantum
  variational measurement in the next generation gravitational-wave
  detectors}},}\ }\href@noop {} {\bibfield  {journal} {\bibinfo  {journal}
  {Physical Review D}\ }\textbf {\bibinfo {volume} {76}},\ \bibinfo {pages}
  {102002} (\bibinfo {year} {2007})}\BibitemShut {NoStop}%
\bibitem [{\citenamefont {Rempe}\ \emph {et~al.}(1992)\citenamefont {Rempe},
  \citenamefont {Thompson}, \citenamefont {Kimble},\ and\ \citenamefont
  {Lalezari}}]{Rempe1992}%
  \BibitemOpen
  \bibfield  {author} {\bibinfo {author} {\bibfnamefont {G.}~\bibnamefont
  {Rempe}}, \bibinfo {author} {\bibfnamefont {R.~J.}\ \bibnamefont {Thompson}},
  \bibinfo {author} {\bibfnamefont {H.~J.}\ \bibnamefont {Kimble}}, \ and\
  \bibinfo {author} {\bibfnamefont {R.}~\bibnamefont {Lalezari}},\ }\bibfield
  {title} {\enquote {\bibinfo {title} {{Optics InfoBase: Optics Letters -
  Measurement of ultralow losses in an optical interferometer}},}\ }\href@noop
  {} {\bibfield  {journal} {\bibinfo  {journal} {Optics letters}\ } (\bibinfo
  {year} {1992})}\BibitemShut {NoStop}%
\bibitem [{\citenamefont {Ueda}\ \emph {et~al.}(1996)\citenamefont {Ueda},
  \citenamefont {Uehara}, \citenamefont {Uchisawa}, \citenamefont {Ueda},
  \citenamefont {Sekiguchi}, \citenamefont {Mitake}, \citenamefont {Nakamura},
  \citenamefont {Kitajima},\ and\ \citenamefont {Kataoka}}]{Ueda1996}%
  \BibitemOpen
  \bibfield  {author} {\bibinfo {author} {\bibfnamefont {A.}~\bibnamefont
  {Ueda}}, \bibinfo {author} {\bibfnamefont {N.}~\bibnamefont {Uehara}},
  \bibinfo {author} {\bibfnamefont {K.}~\bibnamefont {Uchisawa}}, \bibinfo
  {author} {\bibfnamefont {K.-i.}\ \bibnamefont {Ueda}}, \bibinfo {author}
  {\bibfnamefont {H.}~\bibnamefont {Sekiguchi}}, \bibinfo {author}
  {\bibfnamefont {T.}~\bibnamefont {Mitake}}, \bibinfo {author} {\bibfnamefont
  {K.}~\bibnamefont {Nakamura}}, \bibinfo {author} {\bibfnamefont
  {N.}~\bibnamefont {Kitajima}}, \ and\ \bibinfo {author} {\bibfnamefont
  {I.}~\bibnamefont {Kataoka}},\ }\bibfield  {title} {\enquote {\bibinfo
  {title} {{Ultra-High Quality Cavity with 1.5 ppm Loss at 1064 nm}},}\
  }\href@noop {} {\bibfield  {journal} {\bibinfo  {journal} {Optical Review}\
  }\textbf {\bibinfo {volume} {3}},\ \bibinfo {pages} {369--372} (\bibinfo
  {year} {1996})}\BibitemShut {NoStop}%
\bibitem [{\citenamefont {Sato}\ \emph {et~al.}(1999)\citenamefont {Sato},
  \citenamefont {Miyoki}, \citenamefont {Ohashi},\ and\ \citenamefont
  {Fujimoto}}]{Sato1999}%
  \BibitemOpen
  \bibfield  {author} {\bibinfo {author} {\bibfnamefont {S.}~\bibnamefont
  {Sato}}, \bibinfo {author} {\bibfnamefont {S.}~\bibnamefont {Miyoki}},
  \bibinfo {author} {\bibfnamefont {M.}~\bibnamefont {Ohashi}}, \ and\ \bibinfo
  {author} {\bibfnamefont {M.~K.}\ \bibnamefont {Fujimoto}},\ }\bibfield
  {title} {\enquote {\bibinfo {title} {{Optics InfoBase: Applied Optics - Loss
  Factors of Mirrors for a gravitational wave Antenna}},}\ }\href@noop {}
  {\bibfield  {journal} {\bibinfo  {journal} {Applied {\ldots}}\ } (\bibinfo
  {year} {1999})}\BibitemShut {NoStop}%
\bibitem [{\citenamefont {Kells}(2007)}]{Kells2007}%
  \BibitemOpen
  \bibfield  {author} {\bibinfo {author} {\bibfnamefont {W.}~\bibnamefont
  {Kells}},\ }\href@noop {} {\enquote {\bibinfo {title} {{Initial LIGO COC Loss
  investigation Summary}},}\ }\bibinfo {type} {Tech. Rep.}\ \bibinfo {number}
  {LIGO-T070051}\ (\bibinfo  {institution} {CALIFORNIA INSTITUTE OF
  TECHNOLOGY},\ \bibinfo {year} {2007})\ \bibinfo {note}
  {https://dcc.ligo.org/LIGO-T070051}\BibitemShut {NoStop}%
\bibitem [{\citenamefont {Battesti}\ \emph {et~al.}(2007)\citenamefont
  {Battesti}, \citenamefont {Pinto Da~Souza}, \citenamefont {Batut},
  \citenamefont {Robilliard}, \citenamefont {Bailly}, \citenamefont {Michel},
  \citenamefont {Nardone}, \citenamefont {Pinard}, \citenamefont {Portugall},
  \citenamefont {Tr{\'e}nec}, \citenamefont {Mackowski}, \citenamefont
  {Rikken}, \citenamefont {Vigu{\'e}},\ and\ \citenamefont
  {Rizzo}}]{Battesti2007}%
  \BibitemOpen
  \bibfield  {author} {\bibinfo {author} {\bibfnamefont {R.}~\bibnamefont
  {Battesti}}, \bibinfo {author} {\bibfnamefont {B.}~\bibnamefont {Pinto
  Da~Souza}}, \bibinfo {author} {\bibfnamefont {S.}~\bibnamefont {Batut}},
  \bibinfo {author} {\bibfnamefont {C.}~\bibnamefont {Robilliard}}, \bibinfo
  {author} {\bibfnamefont {G.}~\bibnamefont {Bailly}}, \bibinfo {author}
  {\bibfnamefont {C.}~\bibnamefont {Michel}}, \bibinfo {author} {\bibfnamefont
  {M.}~\bibnamefont {Nardone}}, \bibinfo {author} {\bibfnamefont
  {L.}~\bibnamefont {Pinard}}, \bibinfo {author} {\bibfnamefont
  {O.}~\bibnamefont {Portugall}}, \bibinfo {author} {\bibfnamefont
  {G.}~\bibnamefont {Tr{\'e}nec}}, \bibinfo {author} {\bibfnamefont {J.~M.}\
  \bibnamefont {Mackowski}}, \bibinfo {author} {\bibfnamefont {G.~L. J.~A.}\
  \bibnamefont {Rikken}}, \bibinfo {author} {\bibfnamefont {J.}~\bibnamefont
  {Vigu{\'e}}}, \ and\ \bibinfo {author} {\bibfnamefont {C.}~\bibnamefont
  {Rizzo}},\ }\bibfield  {title} {\enquote {\bibinfo {title} {{The BMV
  experiment: a novel apparatus to study the propagation of light in a
  transverse magnetic field}},}\ }\href@noop {} {\bibfield  {journal} {\bibinfo
   {journal} {The European Physical Journal D}\ }\textbf {\bibinfo {volume}
  {46}},\ \bibinfo {pages} {323--333} (\bibinfo {year} {2007})}\BibitemShut
  {NoStop}%
\bibitem [{\citenamefont {Maga{\~n}a-Sandoval}\ \emph
  {et~al.}(2012)\citenamefont {Maga{\~n}a-Sandoval}, \citenamefont {Adhikari},
  \citenamefont {Frolov}, \citenamefont {Harms}, \citenamefont {Lee},
  \citenamefont {Sankar}, \citenamefont {Saulson},\ and\ \citenamefont
  {Smith}}]{Magana2012}%
  \BibitemOpen
  \bibfield  {author} {\bibinfo {author} {\bibfnamefont {F.}~\bibnamefont
  {Maga{\~n}a-Sandoval}}, \bibinfo {author} {\bibfnamefont {R.~X.}\
  \bibnamefont {Adhikari}}, \bibinfo {author} {\bibfnamefont {V.}~\bibnamefont
  {Frolov}}, \bibinfo {author} {\bibfnamefont {J.}~\bibnamefont {Harms}},
  \bibinfo {author} {\bibfnamefont {J.}~\bibnamefont {Lee}}, \bibinfo {author}
  {\bibfnamefont {S.}~\bibnamefont {Sankar}}, \bibinfo {author} {\bibfnamefont
  {P.~R.}\ \bibnamefont {Saulson}}, \ and\ \bibinfo {author} {\bibfnamefont
  {J.~R.}\ \bibnamefont {Smith}},\ }\bibfield  {title} {\enquote {\bibinfo
  {title} {Large-angle scattered light measurements for quantum-noise filter
  cavity design studies},}\ }\href@noop {} {\bibfield  {journal} {\bibinfo
  {journal} {JOSA A}\ }\textbf {\bibinfo {volume} {29}},\ \bibinfo {pages}
  {1722--1727} (\bibinfo {year} {2012})}\BibitemShut {NoStop}%
\bibitem [{\citenamefont {Buonanno}\ and\ \citenamefont
  {Chen}(2001)}]{Buonanno2001}%
  \BibitemOpen
  \bibfield  {author} {\bibinfo {author} {\bibfnamefont {A.}~\bibnamefont
  {Buonanno}}\ and\ \bibinfo {author} {\bibfnamefont {Y.}~\bibnamefont
  {Chen}},\ }\bibfield  {title} {\enquote {\bibinfo {title} {Quantum noise in
  second generation, signal-recycled laser interferometric gravitational-wave
  detectors},}\ }\href {\doibase 10.1103/PhysRevD.64.042006} {\bibfield
  {journal} {\bibinfo  {journal} {Phys. Rev. D}\ }\textbf {\bibinfo {volume}
  {64}},\ \bibinfo {pages} {042006} (\bibinfo {year} {2001})}\BibitemShut
  {NoStop}%
\bibitem [{\citenamefont {Caves}\ and\ \citenamefont
  {Schumaker}(1985)}]{Caves1985}%
  \BibitemOpen
  \bibfield  {author} {\bibinfo {author} {\bibfnamefont {C.~M.}\ \bibnamefont
  {Caves}}\ and\ \bibinfo {author} {\bibfnamefont {B.~L.}\ \bibnamefont
  {Schumaker}},\ }\bibfield  {title} {\enquote {\bibinfo {title} {{New
  formalism for two-photon quantum optics. I. Quadrature phases and squeezed
  states}},}\ }\href@noop {} {\bibfield  {journal} {\bibinfo  {journal}
  {Physical Review A}\ }\textbf {\bibinfo {volume} {31}},\ \bibinfo {pages}
  {3068} (\bibinfo {year} {1985})}\BibitemShut {NoStop}%
\bibitem [{\citenamefont {Yuen}(1976)}]{Yuen1976}%
  \BibitemOpen
  \bibfield  {author} {\bibinfo {author} {\bibfnamefont {H.}~\bibnamefont
  {Yuen}},\ }\bibfield  {title} {\enquote {\bibinfo {title} {{Two photon
  coherent states of the radiation field}},}\ }\href@noop {} {\bibfield
  {journal} {\bibinfo  {journal} {Phys.Rev.}\ }\textbf {\bibinfo {volume}
  {A13}},\ \bibinfo {pages} {2226--2243} (\bibinfo {year} {1976})}\BibitemShut
  {NoStop}%
\bibitem [{\citenamefont {Corbitt}, \citenamefont {Chen},\ and\ \citenamefont
  {Mavalvala}(2005)}]{Corbitt2005}%
  \BibitemOpen
  \bibfield  {author} {\bibinfo {author} {\bibfnamefont {T.}~\bibnamefont
  {Corbitt}}, \bibinfo {author} {\bibfnamefont {Y.}~\bibnamefont {Chen}}, \
  and\ \bibinfo {author} {\bibfnamefont {N.}~\bibnamefont {Mavalvala}},\
  }\bibfield  {title} {\enquote {\bibinfo {title} {{Mathematical framework for
  simulation of quantum fields in complex interferometers using the two-photon
  formalism}},}\ }\href@noop {} {\bibfield  {journal} {\bibinfo  {journal}
  {Physical Review A}\ }\textbf {\bibinfo {volume} {72}} (\bibinfo {year}
  {2005})}\BibitemShut {NoStop}%
\end{thebibliography}%

%%%%%%%%%%%%%%%%%%%%%%%%%%%%%%%%%%%%%%%%
%%%%%%%%%%%%%%%%%%%%%%%%%%%%%%%%%%%%%%%%
\end{document}